\documentclass[11pt]{article}

\usepackage[dvips]{graphicx}
\usepackage{cite}

\begin{document}
\twocolumn [
\title{\textsc{On Friedmann integrals coincidence}}
\author{O. B. Karpov\\
        \emph{Moscow State Mining University,}\\
        \emph{119991, Moscow, Russia}}
\maketitle \vspace{-7ex}
\renewcommand{\abstractname}{ }\abstractname
\begin{abstract}
\textsf{Expansion of the two-component universe with arbitrary
spatial curvature has been considered. It has been shown that the
Friedmann integrals of the almost flat universe do not coincide.}
\end{abstract}\\
\vspace{7ex}]

    As it is known, luminosity distance--redshift relationship for
the type Ia supernovae reveals the presence of an acceleration of
the cosmological expansion \cite{1, 2}. This is possible in
existence of a cosmological repulsion, which is described usually
by $\Lambda$-term in the Einstein equations or equivalently by the
presence of a vacuum-like medium with a negative pressure. The
expansion becomes accelerated when the doubled vacuum density
exceeds the decreasing matter density. These two components are
described by two evolution constants being named Friedmann
integrals. In recent works \cite{3, 4, 5} it is claimed that the
known cosmic coincidences \cite{6} have to be supplemented by a
Friedmann integrals coincidence, and various reasons and
consequences of this coincidence are investigated. In the present
work it is shown that the Friedmann integrals of our almost flat
Universe do not and cannot coincide.

    The two-component universe dynamics is described by the Einstein
field equation for the expansion factor $a$,
\begin{equation}\label{eq:1}
\dot{a}^{2}=\frac{A_{m}}{a}+\frac{a^{2}}{A_{v}^{2}}-k,
\end{equation}
where $k=1,\, 0,\, -1$ for positive, zero and negative spatial
curvature in comoving with the cosmological expansion frame. In
the case $k=\pm 1$ factor $a$ is the space curvature radius, and
for the flat universe $k=0$ factor $a$ is defined up to arbitrary
scale transformation. Here and below convention $c=1$ is used.

    The Friedmann integrals $A_{m}$ and $A_{v}$
\begin{equation}\label{eq:2}
A_{m}=\frac{8\pi}{3}G\rho_{m}a^{3}, \quad
A_{v}^{-2}=\frac{8\pi}{3}G\rho_{v}=\frac{\Lambda}{3}
\end{equation}
determine the universe evolution. Here $\rho_{m}$ is the dust-like
matter energy density including a dark matter, $\rho_{v}$ is the
vacuum energy density, $\Lambda$ is the cosmological constant, $G$
is the gravitational constant. Define
\[
U_{m}=-\frac{A_{m}}{a},\quad U_{v}=-\frac{a^{2}}{A_{v}^{2}}, \quad
U=U_{m}+U_{v}.
\]
Graphs of the functions $U_{m}(a)$, $U_{v}(a)$, $U(a)$ are plotted
in Fig. 1.
\begin{figure}[h]
    \centering
    \includegraphics[width=0.45\textwidth]{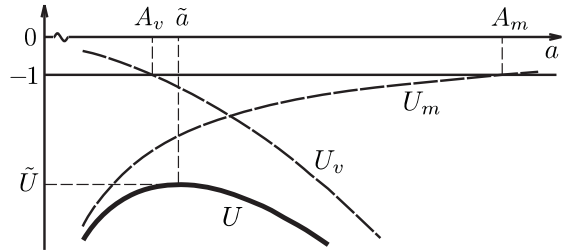}
    \caption{Potential barrier and Friedmann integrals.}
\end{figure}
When $k=1$, the Friedmann integral $A_{m}$ is a maximum value of
the curvature radius in the standard model $\Lambda=0$, the
integral $A_{v}$ equals to an initial value $a$ in the de Sitter
universe $\rho_{m}=0$. At the point $a=\tilde{a}$,
\begin{equation}\label{eq:3}
\tilde{a}^{3}=\frac{1}{2}A_{m}A_{v}^{2}
\end{equation}
the maximum $\tilde{U}$ of the potential barrier $U(a)$ is
disposed,
\begin{equation}\label{eq:4}
\tilde{U} = U(\tilde{a}) = -\left(\frac{\alpha
A_{m}}{A_{v}}\right)^{2/3},
\end{equation}
where $\alpha=3\sqrt{3}/2$. At this point the balance of matter
gravitation and vacuum antigravitation is achieved, $\rho_{m} =
2\rho_{v}$, and the deceleration parameter $q$ changes the sign:
\begin{equation}\label{eq:5}
q = -\frac{\ddot{a}a}{\dot{a}^{2}} =
\frac{\tilde{a}^{3}-a^{3}}{a^{3}-aA_{v}^{2}+2\tilde{a}^{3}}.
\end{equation}
Density parameters $\Omega_{m}=\rho_{m}/\rho_{c}$,
$\Omega_{v}=\rho_{v}/\rho_{c}$, where  $\rho_{c}$ is the critical
density, $\rho_{c}=3H^{2}/8\pi G$, $H=\dot{a}/a$ is the Hubble
constant. The Einstein field equation (\ref{eq:1}) connects the
density parameters and the Friedmann integrals by the correlation
\begin{equation}\label{eq:6}
\left(\Omega - 1\right)^{3} =
\left(\frac{A_{v}}{A_{m}}\right)^{2}\Omega_{m}^{2}\Omega_{v}k^{3},
\end{equation}
where $\Omega$ is the total density parameter, $\Omega=1$ for the
flat universe; in the model (\ref{eq:1}) $\Omega = \Omega_{m} +
\Omega_{v}$.

    As it is clear from Fig. 1, the Friedmann integrals coincide
if the curves $U_{m}(a)$ and $U_{v}(a)$ intersection point is on
the level $U=-1$. If the level line $U=-1$ passes through the
potential barrier summit then $A_{v}=\alpha A_{m}$. In the case
$|\tilde{U}|>1$ i.e. $A_{v}>\alpha A_{m}$ the closed universe
cannot overcome the potential barrier and do not reach the
accelerated expansion state \cite{7}.

    According to data adduced in \cite{3, 4, 5}, $A_{v}$
does not differ from $A_{m}$ more than one order of magnitude,
$A_{v}$ being larger than $A_{m}$, and the expansion factor now
$a_{0}$ coincides with $A_{v}$. For our almost flat Universe
$\Omega_{v} \simeq 0.7$, $\Omega_{m} \simeq 0.3$. As it is clear
from the correlation (\ref{eq:6}), irrespective of the curvature
sign the Friedmann integrals ratio $A_{v}/A_{m}$ is connected with
the degree of the universe spatial flatness: $\Omega\rightarrow 1
\Leftrightarrow A_{v}/A_{m} \rightarrow 0$, and the barrier summit
$\tilde{U}$ (\ref{eq:4}) goes down from the level $U=-1$. Because
of that for the almost flat universe $a_{0}$ and $A_{v}$ do not
coincide either:
\begin{equation}\label{eq:7}
\left(\frac{a_{0}}{A_{v}}\right)^{3} =
\frac{\Omega_{v}}{\Omega_{m}} \frac{A_{m}}{A_{v}} \gg 1,
\end{equation}
and $a/A_{v}\rightarrow\infty$ when $\Omega\rightarrow 1$. For all
this
\begin{equation}\label{eq:8}
\left(\frac{a_{0}}{A_{m}}\right)^{3} =
\frac{\Omega_{v}}{\Omega_{m}} \frac{A_{v}}{A_{m}} \ll 1.
\end{equation}
Last data on anisotropy of the cosmic microwave background
restrict spatial curvature of the Universe \cite{8}
$|\Omega-1|\leq 0.03$. According to (\ref{eq:6}) it corresponds
$A_{v}/A_{m}\leq 2\cdot10^{-2}$. Emphasize that the foregoing
statements on the value of the Friedmann integrals ratio are
correct for $k=\pm 1$. A flat space is the limit of the curved one
when $A_{v}/A_{m}\rightarrow 0$. The latter is also clear from the
fact that $A_{v}$ does not depend on the curvature radius and
$A_{m}\propto a^{3}$ (\ref{eq:2}). For the formally flat universe,
when $k=0$, the equation (\ref{eq:6}) turns into an identity and
in equation (\ref{eq:1}) the parameter $a$ becomes an arbitrary
scale factor. Redefining the latter one can make any value of
$A_{m}$ and the ratio of the Friedmann integrals loses the sense.


\begin{thebibliography}{8}
\bibitem{1} A.G. Riess \emph{et al.}, Astron. J. \textbf{116} 1009
            (1998), astro-ph/9805201.
\bibitem{2} S. Perlmutter \emph{et al.}, Astrophys. J. \textbf{517} 565
            (1999), astro-ph/9812133.
\bibitem{3} A.D. Chernin, Uspekhi Fisicheskikh Nauk \textbf{171} 153 (2001).
\bibitem{4} A.D. Chernin, astro-ph/0101532.
\bibitem{5} A.D. Chernin, New Astron. \textbf{7} 113 (2002),
            astro-ph/0107071.
\bibitem{6} J. Garriga, A. Vilenkin, Phys. Rev. D \textbf{64}
            023517 (2001), hep-th/0011262.
\bibitem{7} O.B. Karpov, gr-qc/0301039.
\bibitem{8} A. Benoit \emph{et al.}, astro-ph/0210306.
\end{thebibliography}
\end{document}